\documentclass[aps,prb]{revtex4}
\usepackage{graphicx}

\newcommand{\be}{\begin{equation}}
\newcommand{\ee}{\end{equation}}
\newcommand{\ba}{\begin{array}}
\newcommand{\ea}{\end{array}}
\newcommand{\bea}{\begin{eqnarray}}
\newcommand{\eea}{\end{eqnarray}}

\begin{document}

\textheight=23cm
\topmargin=0cm

\title{Hysteresis in Ferromagnetic Random Field Ising Model with an
Arbitrary Initial State }

\author{Prabodh Shukla } 
\email{shukla@nehu.ac.in} 
\author{R Kharwanlang} 
\affiliation{ Physics Department \\ North Eastern Hill University
\\Shillong-793 022,India.}

\begin{abstract}

We present exact expressions for hysteresis loops in the ferromagnetic
random field Ising model in the limit of zero temperature and zero driving
frequency for an arbitrary initial state of the model on a Bethe lattice.
This work extends earlier results that were restricted to an initial state
with all spins pointing parallel to each other.

\end{abstract}

\maketitle

\section{Introduction} 

Zero temperature Glauber dynamics of a driven ferromagnetic random field
Ising model (RFIM) provides an interesting caricature of hysteresis,
Barkhausen noise, and return point memory in several complex systems
~\cite{sethna}.  It also provides a simple example of non-equilibrium
critical behavior along with scaling and universality commonly associated
with equilibrium critical phenomena. In view of these features, the model
has been studied extensively by numerical simulations on hypercubic
lattices, and by exact solution in one dimension, and on a Bethe lattice
in special cases ~\cite{shukla1,dhar,sabha1,shukla2, sabha2}. Results on
the Bethe lattice show an unexpected dependence of non-equilibrium
critical behavior on the coordination number of the lattice. The
non-equilibrium critical point occurs only if $z>3$, where $z$ is the
coordination number of the lattice. These and other results have been
obtained in the limit of zero temperature and zero driving frequency,
starting from a state of saturated magnetization, i.e. when all spins are
initially pointing parallel to each other. In the present paper, we extend
the earlier results by obtaining an analytic expression for the hysteresis
loop starting from an arbitrary initial state.

\section{The Model}

Ising spins $S_{i}$ ($S_i= \pm 1$ ) are located on a lattice at sites
labeled by an integer i = 1 to N. Nearest neighbor spins interact via a
ferromagnetic interaction J (J$>$0). A site-dependent quenched random
field $h_i$ acts on $S_i$ in addition to a uniform external field $h$ that
acts on all spins. The fields $\{h_i\}$ are independent identically
distributed random variables drawn from a continuous probability
distribution $\phi(h_{i})$. The Hamiltonian of the system is

\be H=-J \sum_{i,j} S_{i}S_{j} - \sum_{i} h_{i}S_{i} - h \sum_{i} S_{i}
=-\sum_{i} f_iS_i, \ee

where $f_i$ is the net field acting on the spin $S_i$: $ f_{i}=J \sum_{j}
S_{j} + h_{i} + h $. The system evolves through single-spin-flip Glauber
dynamics at zero temperature that lowers the energy of the system by
aligning spins along the net field at their respective sites, i.e. $S_{i}$
is flipped if it does not have the same sign as $f_{i}$. The dynamics is
an iterative process because flipping of a spin alters the net field on
its neighbors, and may necessitate the flipping of neighbors and their
neighbors. In practice, the dynamics is implemented by the following rule:
choose a spin at random, and flip it only if it is not aligned along the
net field at its site.  Repeat the process till all spins are aligned
along the net fields at their respective sites.

\section{Hysteresis on a Bethe Lattice}

A Bethe lattice is an infinite-size branching tree of coordination number
$z$. We consider a random configuration of spins on this lattice such that
the spin on any given site is up with probability $f$ and down with
probability $1-f$. Two special values of $f$ equal to zero and unity that
correspond to an initial state with all spins parallel to each other have
been examined earlier in the study of hysteresis on a Bethe lattice. The
purpose of the present paper is to extend the analytic results obtained
earlier to all values of $f$ in the range $0\le f \le1$. As the external
field is slowly ramped up from $-\infty$ to $\infty$, we ask what fraction
of spins are up at an intermediate applied field $h$. We choose a site at
random in the deep interior of the tree, and designate it as the central
site. We need to calculate the probability $p(h)$ that the central site is
up when the applied field has been increased from $-\infty$ to $h$. The
fraction $f$ sets the lower bound on $p(h)$ because spins that are up in
the initial state must remain up in an increasing field. Indeed, the
fraction $f$ of spins that are up in the initial state may be considered
"quenched"  because they do not form dynamical variables of the problem in
an increasing applied field. Thus the problem reduces to finding the
probability that the central site is up given that it was down initially.
In the following, we calculate the complementary probability $q(h)=1-p(h)$
that the central site is down when the system is relaxed at field $h$. The
magnetization per site is given by $m(h)=1-2 q(h)$.

Each nearest neighbor of the central site forms the vertex of an infinite
sub-tree. The sub-trees do not interact with each other except through the
central site. Therefore the evolution on $z$ sub-trees meeting at the
central site is independent of each other as long as the central site does
not flip up from its initial state. In ferromagnetic dynamics the same
final stable state is obtained irrespective of the order in which we relax
the central site and its nearest neighbors. This is known as the Abelian
property of the ferromagnetic dynamics. We choose to relax the central
site after its neighbors have been relaxed. Let $Q^{*}(h)$ denote the
conditional probability that a nearest neighbor of the central site is
down after it has been relaxed at field $h$ given that the central site is
is held down at $h$. The central site is relaxed after its nearest
neighbors have been relaxed. The probability that the central site stays
down after it is allowed to relax is given by,

\be q(h)=(1 - f) \sum_{m=0}^{z} \left( \ba{c} {z} \\ {m} \ea
\right)[1-Q^{*}(h)]^{m} [Q^{*}(h)]^{z-m} [1 - p_{m}(h)] \ee

The first factor $(1-f)$ gives the probability that the central site is
initially down, the last factor, $[1-p_{m}(h)]$, gives the probability
that a spin with $m$ nearest neighbors up, and $z-m$ neighbors down does
not have sufficient random quenched field at its site to flip it up at an
applied field $h$.

\be p_{m}(h) = \int_{(z - 2m)J-h}^{\infty} \phi(h_{i}) dh_{i} \mbox{ 
\hspace{1cm} (m=0,\ldots,z)} \ee

The conditional probability $Q^{*}(h)$ that enters equation (2) is 
given by the fixed point of the following recursion relation.

\be Q^{*}=\lim_{n \rightarrow \infty} Q^{n}; \mbox{ }Q^{n}=(1-f)
\sum_{m=0}^{z-1} \left( \ba{c}{z-1} \\ {m} \ea \right)  [1 - Q^{n-1}]^{m}
[Q^{n-1}]^{z-1-m} (1-p_m) \ee

The rationale for the recursion relation (4) is as follows. Each nearest
neighbor of the central site forms the vertex of a sub-tree. Let the
central site and one of its nearest neighbors be at a distance of $n+1$
and $n$ steps respectively from the boundary of the corresponding
sub-tree. $Q^n$ denotes the conditional probability that the vertex is
down when relaxed at field $h$ given that the central site is down before
the vertex is relaxed. In our scheme, the sites at the boundary of the
sub-tree are relaxed first, then the sites located one step above the
boundary and so on. The recursion relation works one step at a time from
the boundary to the vertex. Only non-quenched sites are relaxed. The
probability that the vertex is not quenched is equal to $1-f$. This
accounts for the first factor on the right side of the recursion relation.
Given that the central site is down, the vertex has $z-1$ neighbors at
height $n-1$ that than be independently down or up as a result of the
branching structure of the tree. $Q^{n-1}$ is the conditional probability
that any one of the $(z-1)$ neighbors at the lower level is down when
relaxed at $h$ given that the vertex is down; $[1-Q^{n-1}]$ is the
probability that a neighbor at the lower level is up. The up neighbor may
be a quenched or a non-quenched site. The three factors following the
summation sign in equation (4) give the probability that $m$ out of
$(z-1)$ sites are up, and the rest are down. The last factor $(1-p_m)$
gives the probability that the vertex remains down when relaxed. The
recursion relation (4) is similar to the one obtained earlier in the case
when the initial state has all spins pointing parallel to it, and reduces
to the earlier result if $f=0$, as it should. However, for $0 \le f \le 1$
the lattice is punctuated by quenched sites, and we relax only the
non-quenched sites. The relaxed state of the vertex is determined by the
longest path of "non-quenched" sites connected to it. The longest possible
path is the one that connects the boundary to the vertex. As, $n
\rightarrow \infty$, the recursion relation iterates to the fixed point
solution $Q^{*}(h)$ which incorporates in it all shorter paths with
suitable weights.

If the applied field is reversed before completing the lower half of the
major loop, a minor hysteresis loop is generated. First reversal of the
field generates the upper half of the minor loop, and a second reversal
generates the lower half.  When the field on second reversal reaches the
point where the first reversal was made, the lower half of the minor loop
meets the starting point of the upper half. In other words, the minor loop
closes upon itself at the point it started. This property of the RFIM is
known as return point memory. The analytic calculation of minor loop is
somewhat more tedious than that of the major loop. Consider the upper half
of the minor loop. Suppose the applied field is reversed from $h$ to
$h^{\prime}$ ($h^{\prime} \le h$). By assumption, the quenched sites do
not turn down in the reversed field. We need to calculate the probability
that a non-quenched site that was up at $h$ turns down at $h^{\prime}$.
When a non-quenched site $i$ turns up, the field on its nearest neighbors
increases by an amount $2J$. The increased field may cause some
non-quenched neighbors to turn up. Each neighbor that turns up after site
$i$, increases the field on site $i$ by an amount $2J$.  Therefore, in the
reversing field site $i$ can turn down only after all neighbors which
turned up after it have turned down. The probability $D^{*}(h^{\prime})$
that a non-quenched nearest neighbor of site $i$ that was down before site
$i$ turned up is again down at $h^{\prime}$ is determined by the fixed
point of the following recursion relation,

\bea D^{n}(h^{\prime})= (1-f) \sum_{m=0}^{z-1} \left( \ba{c}{z-1} \\ {m}
\ea \right) [1-Q^{*}(h)]^{m}[Q^{*}(h)]^{z-1-m} \left[1-p_{m+1}(h)\right] &
\nonumber \\ + (1-f) \sum_{m=0}^{z-1} \left( \ba{c} {z-1} \\ {m} \ea
\right) [1-Q^{*}(h)]^{m}[D^{n-1}(h^{\prime})]^{z-1-m} \left[p_{m+1}(h) -
p_{m+1}(h^{\prime})\right] \eea

Given a site $i$ that is up at $h$, the first sum above gives the
conditional probability that a non-quenched nearest neighbor of site $i$
remains down at $h$ after site $i$ has turned up. The second sum takes
into account the situation that the nearest neighbor in question turns up
at $h$ after site $i$ turns up but turns down at $h^{\prime}$.

The fraction of non-quenched sites that turn down at $h^{\prime}$ is given
by,

\be q^{\prime}(h^{\prime})= (1-f) \sum_{m=0}^{z} \left( \ba{c} {z} \\ {m}
\ea \right) [1-Q^{*}(h)]^{m}[D^{*}(h^{\prime})]^{z-m}
\left[p_{m}(h)-p_{m}(h^{\prime})\right] \ee

Consequently, for $h-2J \le h^{\prime} \le h$, the magnetization on the
upper return loop is given by,

\be m^{\prime}(h^{\prime})=1-2 [q(h)+q^{\prime}(h^{\prime})] \ee

At $h^{\prime}=h-2J$, all neighbors of the central site that flipped up
because the central site flipped up at $h$ have flipped down, and the
neighborhood of the central site is stable with the central site in the up
position. As the applied field decreases further, the central site 
must turn down before any of its nearest neighbors. This means that at 
field $h-2J$ the system arrives at some point on the upper half of the 
major hysteresis loop, and moves on it upon further decrease in the 
applied field. Thus the magnetization for $h^{\prime} < h-2J$ is given by 
the formula,
\be m^{\prime}(h^{\prime})=1-2 \tilde{q}(h^{\prime}), \ee
where,
\be \tilde{q}(h^\prime)=(1 - f) \sum_{m=0}^{z} \left( \ba{c} {z} \\ {m}
\ea \right)[1-\tilde{Q}^{*}(h^\prime)]^{m} [\tilde{Q}^{*}(h^\prime)]^{z-m}
[1 - p_{m}(h^\prime)], \ee

and $\tilde{Q}^{*}(h^\prime)$ is given by the fixed point of the following
recursion relation.
\be \tilde{Q}^{n}=(1-f) \sum_{m=0}^{z-1} \left( \ba{c}{z-1} \\ {m} \ea
\right)  [1 - \tilde{Q}^{n-1}]^{m} [\tilde{Q}^{n-1}]^{z-1-m} (1-p_{m+1})  
\ee

The lower half of the return loop is obtained by reversing the field from
$h^{\prime}$ to $h^{\prime\prime}$ ($h^{\prime\prime} > h^{\prime}$). If
$h^{\prime}<h-2J$, the lower half of the minor loop starts from the major
loop, and therefore it is related by symmetry to the upper half of the
return loop that has been obtained above. We only need to consider the
case $h^{\prime} \ge h-2J$. In this case, the magnetization on the lower
half of the return loop may be written as, \be
m^{\prime\prime}(h^{\prime\prime})=1 -2 [q(h)+q^{\prime}(h^{\prime}) -
p^{\prime\prime}(h^{\prime\prime})], \ee where
$p^{\prime\prime}(h^{\prime\prime})$ is the probability that a
non-quenched site that is up at field $h$ and down at field $h^{\prime}$,
turns up again at $h^{\prime\prime}$.

\be p^{\prime\prime}(h^{\prime\prime})=(1-f) \sum_{m=0}^{z} \left( \ba{c}
{z} \\ {m} \ea \right) [U^{*}(h^{\prime\prime})]^{m}
[D^{*}(h^{\prime})]^{z-m}
\left[p_{m}(h^{\prime\prime})-p_{m}(h^{\prime})\right] \ee

Here $U^{*}(h^{\prime\prime})$ is the conditional probability that
a nearest neighbor of a site $i$ turns up before site $i$ turns up
on the lower return loop. It is determined by the fixed point of the 
equation,

\bea U^{n}(h^{\prime\prime})= 1 - Q^{*}(h) - (1-f) \sum_{m=0}^{z-1} \left(
\ba{c} {z-1} \\ {m} \ea \right) [1-Q^{*}(h)]^{m}
[D^{*}(h^{\prime})]^{z-1-m} [p_{m}(h)-p_{m}(h^{\prime})] & \nonumber \\ +
(1-f) \sum_{m=0}^{z-1} \left( \ba{c} {z-1} \\ {m} \ea \right)  
[U^{n-1}(h^{\prime\prime})]^{m}[D^{*}(h^{\prime})]^{z-1-m}
\left[p_{m}(h^{\prime\prime}) - p_{m}(h^{\prime})\right] \eea

The rationale behind equation (13) is similar to the one behind equation
(5). Given that a site $i$ is down at $h^{\prime}$, the first three terms
on the right hand side account for the probability that a nearest neighbor
of site $i$ is up at $h^{\prime\prime} \ge h^{\prime}$. Note that the
neighbor in question must have been up at $h$ in order to be up at
$h^{\prime}$, and if it it is already up at $h^{\prime}$ then it will
remain up on the entire lower half of the return loop, i.e. at
$h^{\prime\prime} \ge h^{\prime}$. The last term gives the probability
that the neighboring site was down at $h^{\prime}$, but turned up on the
lower return loop before site $i$ turned up. It can be verified that the
lower return loop meets the lower major loop at $h^{\prime\prime}=h$ and
merges with it for $h^{\prime\prime} > h$, thus proving the property of
return point memory.

The method of calculating the minor loop may be extended to obtain a
series of smaller minor loops nested within the minor loop obtained above.  
The key point is that whenever the applied field is reversed, a site $i$
may flip only after all neighbors of site $i$ which flipped in response to
the flipping of site $i$ on the immediately preceding sector have flipped
back. The neighbors of site $i$ which did not flip on the preceding sector
in response to the flipping of site $i$ do not flip in the reversed field
before site $i$ has flipped. We have obtained above expression for the
return loop when the applied field is reversed from $h_{ext}=h$ on the
lower major loop to $h_{ext}=h^{\prime}$ ($h-2J \le h^{\prime} \le h$),
and reversed again from $h_{ext}=h^{\prime}$ to $h_{ext}=h^{\prime\prime}$
($h^{\prime\prime} \le h$). When the applied field is reversed a third
time from $h^{\prime\prime}$ to $h^{\prime\prime\prime}$
($h^{\prime\prime\prime}< h^{\prime\prime}$), expressions for the
magnetization on the nested return loop follow the same structure as the
one on the trajectory from $h$ to $h^{\prime}$. Qualitatively, the role of
$P^{*}$ on the first leg ($h$ to $h^{\prime}$) is taken up by $U^{*}$ on
the third leg ($h^{\prime\prime}$ to $h^{\prime\prime\prime}$) of the
nested return loop.

\section{Comparison of Theory with Simulations}

A branching tree has large surface effects. Special care has to be taken
to eliminate surface effects in theory and simulations before the two can
be compared. We eliminate surface effects somewhat differently in
simulations than in the theoretical formalism. The analytic results on the
Bethe lattice are obtained by taking the infinite-size limit of a
branching tree. Fixed points of recursion relations have the effect of
eliminating surface effects. It is rather difficult to eliminate surface
effects in numerical calculations on branching trees because most of the
sites on a finite tree are on the boundary or close to the boundary. We
therefore perform the numerical simulations of the model on random graphs
of coordination $z$. A random graph of $N$ sites has no surface, but the
price we pay is that it has some loops. However, for almost all sites in
the graph, the local connectivity up to a distance of $
\mbox{log}_{(z-1)}$ N is similar to the one in the deep interior of the
branching tree. Therefore, simulation on a random graph is a very
efficient method of subtracting the surface effects on the corresponding
finite branching tree.

We choose to work with a Gaussian distribution of the random field with
mean value zero, and variance $\sigma^2$ for our numerical work. Our
theoretical expressions are valid for any continuous distribution of the
random field. Figure (1) shows a comparison between theory, and simulation
for z=3, $f=.2$, and $\sigma=1$. A major loop as well as two minor loops
are shown;  the minor loops are obtained by reversing the field at
$h_{1}=.6$ (J=1) on the lower half of the major loop , and making round
trip excursions upto $h_{2}=-1.4$, and $h_{2}=-1$ respectively. Results
from simulations have been superposed on the theoretical curves and are
indistinguishable from it on the scale of the graph.

Hysteresis on a Bethe lattice of coordination $z \ge 4$ is qualitatively
different from the case z=2, and z=3. For $z \ge 4$, there exists a pair
of values of $\sigma$ and $h$ ($\sigma=\sigma_c$, $h=h_{c}$)  that marks a
critical point in the response of the system to the driving field. This
critical point does not exist on lattices with $z=2$, and $3$. The
critical point is marked by the disappearance of a first order jump in the
magnetization in an increasing or decreasing applied field. If $f=0$ and
$\sigma < \sigma_{c}$, the jump in the magnetization occurs at an applied
field $h_{J}$ $>$ J. As $\sigma$ increases to $\sigma_{c}$, $h_{J}$
decreases to J, and the size of the jump reduces to zero.  Thus in the
case $f=0$, $h_c=J$ independent of the value of $z$ as long as $z \ge 4$.
For $z=4$, $\sigma_c=1.78$ approximately. The value of $\sigma_c$
increases with increasing $z$. These features remain qualitatively true
for the enire range $0 \le f \le 1$. When a fraction of spins are
quenched, the system is effectively more disordered and it takes
relatively smaller disorder in the quenched fields to produce the same
effect as in a system with $f=0$. For example, if $f=.2$ the value of
$\sigma_c$ is reduced from 1.78 to 1.2 approximately.  Figure (2) shows
the lower halves of the major hysteresis loop for $z=4$, $f=.2$, and
$\sigma = 1.0, 1.2,$ and $1.4$ respectively. The graph corresponding to
$\sigma=1.4$ is smooth, and the one for $\sigma=1$ has a first order jump
at $h=.334$ approximately.  Results of a simulation have been superimposed
on the theoretical graphs. The agreement is quite satisfactory within
numerical errors.

Figure (3) shows the graph for $z=4$, $f=.2$, and $\sigma=1.0$ in detail.
The magnetization in increasing field jumps from $m \approx .5$ to $m
\approx .97$ at $h_2 \approx .334$.  At an earlier value of the applied
field $h_1 < h_2$ ($h_1 \approx .31$), the line of fixed points of
equation (4) splits into three branches but only one of these branches is
stable in increasing field. The system makes a transition from one stable
branch to another stable branch such that the magnetization varies
smoothly across the field $h_1$. If one starts with an initial value
$Q^0=0$ and iterates recursion relation (4)  to get the fixed point
solution $Q^*$, then $Q^*$ jumps at $h_1 \approx .31$ to a new but
unstable fixed point.  The corresponding magnetization jumps from $m
\approx .25$ to $m \approx .9$ at $h_1 \approx .31$. This is indicated in
figure (3) by a vertical line.  However, the higher value of magnetization
is unstable and it is not observed in simulations. In simulations the
magnetization continues smoothly from $h_1$ to $h_2$ ($h_2 > h_1$), and
jumps up at $h_2$ as indicated in the figure by a broken vertical line.
The magnetization on the lower hysteresis loop from $h_1$ to $h_2$
corresponds to a stable fixed point solution of equation (4) with the
initial condition $Q^0=1$. In the range $h_1 \le h \le h_2$, the fixed
point equation (4)  has three solutions. The magnetizations obtained by
substituting the three fixed point solutions in equation (2) are shown by
an s-shaped curve in figure (3). The middle portion of the s-shaped curve
showing decreasing magnetization in increasing field is unphysical and is
not observed in simulations. The fields $h_1$ and $h_2$ mark the the two
turning points of the s-shaped curve. With increasing $\sigma$ the
unstable segment shrinks, and the s-shaped graph straightens out. At the
critical value $\sigma_c$, the magnetization corresponds to a double root
of the fixed point equations.

\section{Concluding Remarks}

We have obtained analytic expressions for hysteresis loops in the
ferromagnetic random field Ising model on a Bethe lattice in the limit of
zero temperature and zero driving frequency. Our results are applicable to
an arbitrary initial state of the system, and constitute a generalization
of results that were restricted to an initial state with all spins
pointing parallel to each other. We have checked the analytic results
against simulations of the model on random graphs and found the agreement
between theory and simulation to be quite satisfactory. It is remarkable
that a relatively minor modification of the recurrence relations allows us
to pass from an initial state with all spins parallel to each other to a
random initial configuration. This apparent simplicity results from the
use of fixed points of recurrence relations for conditional probabilities.
Let us consider the fixed point solution $Q^*$ for $z=2$:

\be Q^*=\left[ \frac{(1-p_1)}{1-(1-f)(p_1-p_0)}\right] =(1-p_1)
\sum_{m=0}^{\infty} \left[ (1-f)(p_1-p_0) \right]^m \hspace{1cm} (z=2)\ee

Imagine a chain of Ising spins with a fraction $f$ quenched in the up
position, and the rest down at $h=-\infty$. Equation (14) indicates that
given a down spin say at site $i$, the conditional probability that its
nearest neighbor, say on the right side at site $i+1$ is down when relaxed
at field $h$ is equal to the probability that the closest avalanche on the
right side comes to a stop after $m$ steps ($m=1,2,\ldots$).  The
prefactor $(1-p_1)$ gives the probability that the avalanche has stopped,
and the factor after the summation sign gives the probability of an
avalanche of size $m$. The above argument is not entirely transparent. We
have also obtained hysteresis curves for $z=2$ for $f=0$ as well as for
other values of $f$ by an alternate and more tedious method that considers
various probabilities of initiating an avalanche and calculating its
resulting size.  The alternate method gives the same result as the fixed
point method. It shows that a fixed point description of the system
contains information on all fluctuations of the system with appropriate
weights.

Finally, the results presented here can be extended to systems with
impurities and vacancies. In our formalism a fraction $f$ of sites are
quenched in the up position and contribute a constant magnetization (per
site)  equal to $2f-1$. If the constant term is subtracted, and the number
of up nearest neighbors of a site is suitably reduced before that site is
relaxed, the results can be extended to hysteresis in systems that have
quenched random fields, as well as a fraction $f$ of magnetic ions
missing.

\begin{figure}
\begin{center}

\includegraphics[angle=-90,width=16cm]{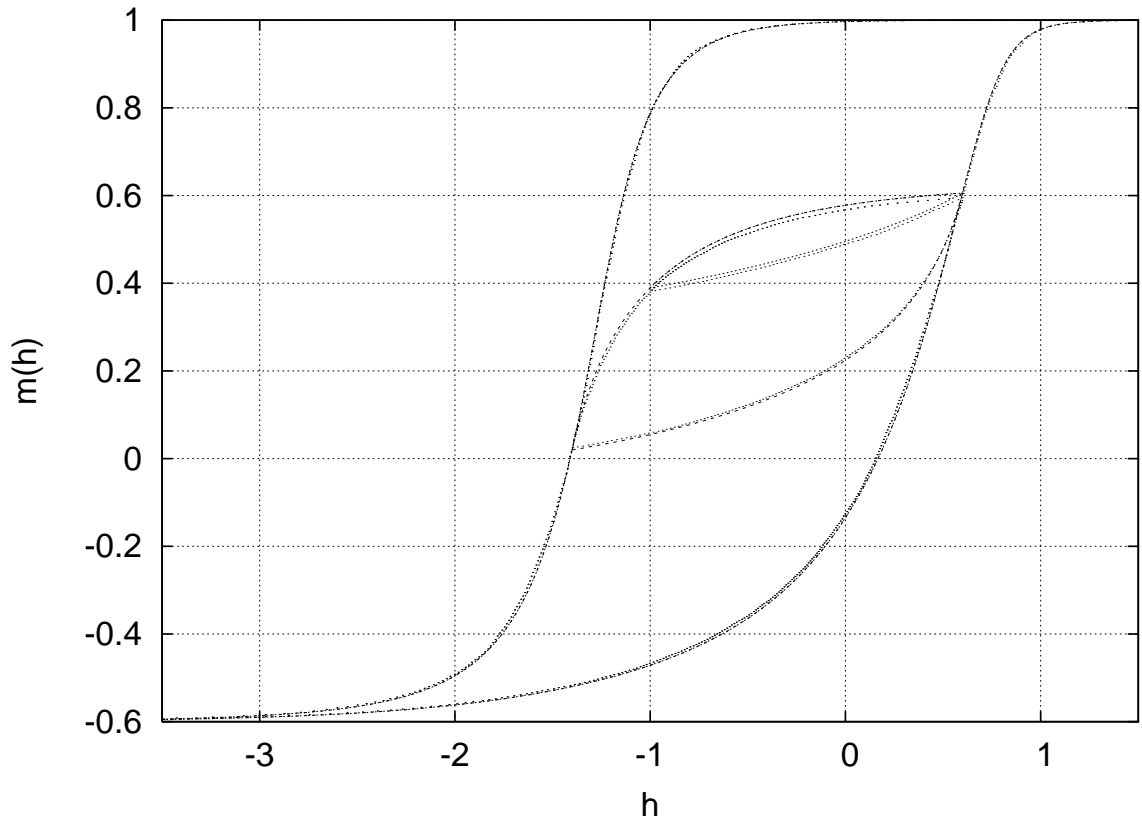}

\caption{Hysteresis on a Bethe lattice with $z=3$, $f=.2$, and $\sigma =
1$. Continuous lines show the theoretical result, and dots mark the
results of numerical simulations. A major hysteresis loop and two minor
loops are shown;  the minor loops are obtained by reversing the field at
$h_{1}=.6$ (J=1) on the lower half of the major loop , and making round
trip excursions upto $h_{2}=-1.4$, and $h_{2}=-1$ respectively. Theory and
simulations are indistinguishable from each other on the scale of the
graph.}

\end{center}
\end{figure}

\begin{figure}
\begin{center}

\includegraphics[angle=-90,width=16cm]{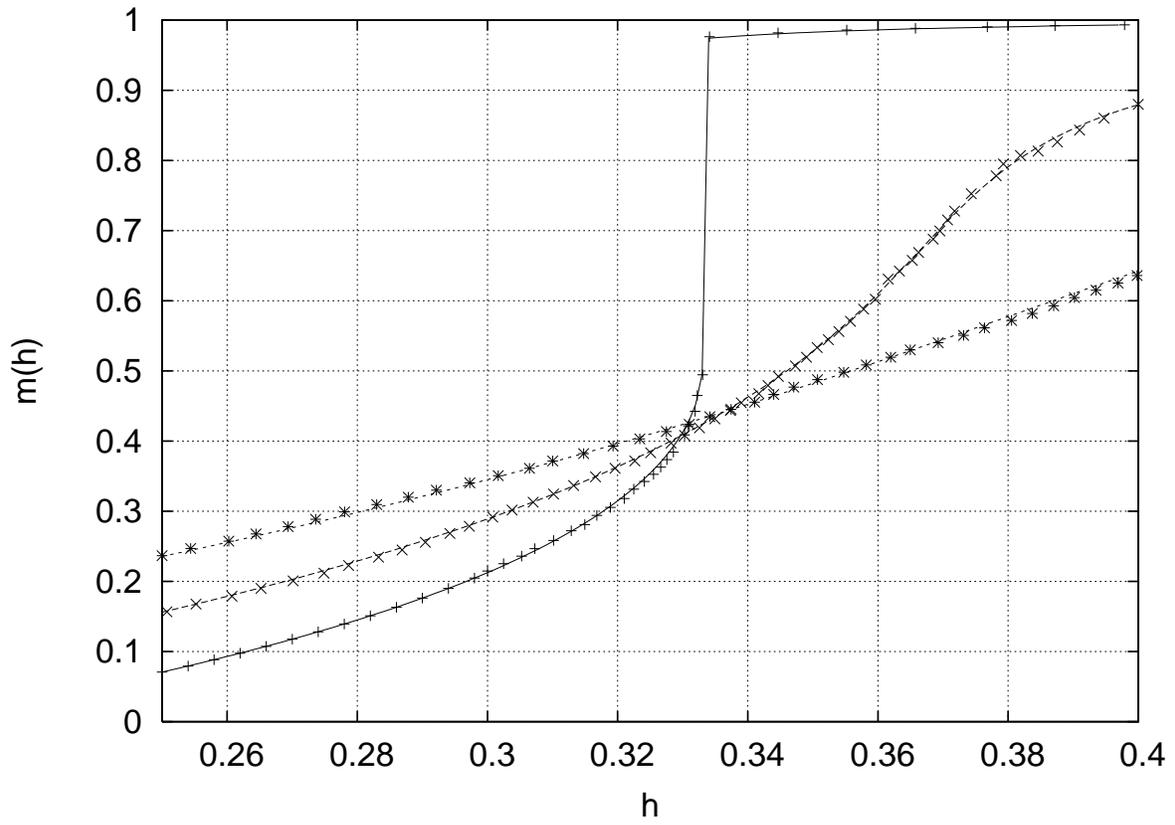}

\caption{Magnetization on a Bethe lattice with $z=4$, and $f=.2$ in an
increasing field. Continuous lines show theoretical results for $\sigma=
1.0$J, $\sigma=1.2$J, and $\sigma=1.4$J . Results from simulations are
superimposed on the corresponding theoretical curves.}

\end{center}
\end{figure}

\begin{figure}
\begin{center}

\includegraphics[angle=-90,width=16cm]{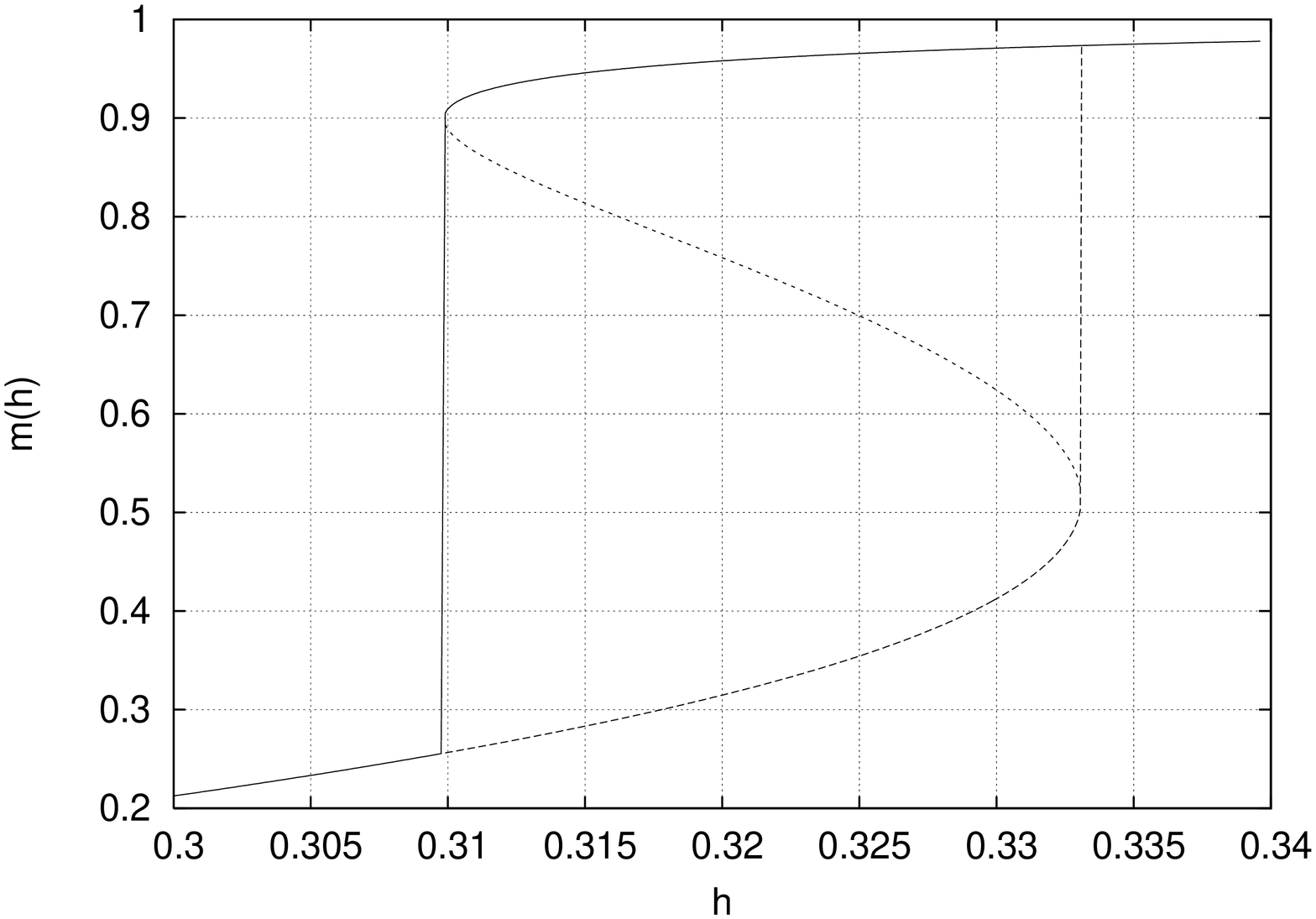}

\caption{The s-shaped magnetization curve ($z=4$, $f=.2$, $\sigma=1.0$J)
showing three fixed point solutions of recursion relations. See text for
details. }

\end{center}
\end{figure}


\begin{thebibliography}{99}

\bibitem{sethna} J P Sethna, K A Dahmen, S Kartha, J A Krumhansl, B W
Roberts, and J D Shore, Phys Rev Lett 70, 3347 (1993); See J P Sethna, K A
Dahmen, and O Perkovic, cond-mat/0406320 for a recent review of
hysteresis, avalanches, and criticality in the ferromagnetic random field
Ising model.

\bibitem{shukla1}P Shukla, Physica A 233, 235 (1996).

\bibitem{dhar}D Dhar, P Shukla, and J P Sethna, J Phys A: Math. Gen.  
30, 5259 (1997).

\bibitem{sabha1} S Sabhapandit, P Shukla, and D Dhar, J Stat Phys 98, 103
(2000).

\bibitem{shukla2} Prabodh Shukla, Phys Rev E 62, 4725 (2000); Phys Rev E
63, 27102 (2001).

\bibitem{sabha2} S Sabhapandit, D Dhar, and P Shukla, Phys Rev Lett 88,
197202 (2002).

\end{thebibliography}
\end{document}